%% file: main.tex
\documentclass[conference]{IEEEtran}
\IEEEoverridecommandlockouts
\usepackage{cite}
\usepackage{amsmath,amssymb,amsfonts}
\usepackage{algorithmic}
\usepackage{graphicx}
\usepackage{float}
\usepackage{subfigure}
\usepackage{placeins}
\usepackage{newfloat}
\usepackage{caption}
\usepackage{authblk}
\DeclareFloatingEnvironment[fileext=frm,placement={!ht},name=PROMPT]{fprompt}

\usepackage[framemethod=TikZ]{mdframed}
\newenvironment{prompt}[1][]{
  \def\mycaption{#1}\begin{fprompt}[tb]
  \begin{mdframed}[roundcorner=10pt,backgroundcolor=blue!10]
}{
  \end{mdframed}\expandafter\caption{\expandafter\mycaption}
  \end{fprompt}
}
\usepackage{textcomp}
\usepackage{xcolor}
\usepackage{colortbl}
\newcommand{\Five}{\cellcolor{green!60} 5}
\newcommand{\Four}{\cellcolor{green!30} 4}
\newcommand{\Three}{\cellcolor{lime!30} 3}
\newcommand{\Two}{\cellcolor{orange!20} 2}
\newcommand{\One}{\cellcolor{red!30} 1}
\newcommand{\Zero}{\cellcolor{red!60} 0}
\usepackage{enumitem}
\usepackage{multirow}
\usepackage{tabularx}
\usepackage{array}
\usepackage{booktabs}
\newcolumntype{L}{>{\raggedright\arraybackslash}X}
\newcolumntype{C}{>{\centering\arraybackslash}X}
\newcolumntype{R}{>{\raggedleft\arraybackslash}X}
\usepackage{hyperref}
\hypersetup{
  colorlinks=true,
  allcolors={blue!50!black},
  pdftitle={Enhancing Finite State Machine Design Automation with Large Language Models and Prompt Engineering Techniques},
  pdfauthor={Qun-Kai Lin, Cheng Hsu, Tian-Sheuan Chang},
  pdfstartview={FitH},
  pdfdisplaydoctitle=true
}
\newcommand{\reflong}[2]{\hyperref[#2]{#1~\ref*{#2}}}

\def\BibTeX{{\rm B\kern-.05em{\sc i\kern-.025em b}\kern-.08em
    T\kern-.1667em\lower.7ex\hbox{E}\kern-.125emX}}
\begin{document}

\bstctlcite{IEEEexample:BSTcontrol}
\title{Enhancing Finite State Machine Design Automation with Large Language Models and Prompt Engineering Techniques}
\author[*]{Qun-Kai Lin}
\author[*]{Cheng Hsu}
\author[ ]{Tian-Sheuan Chang}

\affil[ ]{Department of Electronics and Electrical Engineering, National Yang Ming Chiao Tung University, Hsinchu, Taiwan}
\affil[*]{These authors contributed equally}

\maketitle

\input{abs/abstract_en}

\input{chapters/1-RISCV_SRAM.tex}

\bibliographystyle{IEEEtran}

\bibliography{IEEEabrv,bib/ieeeBSTcontrol,bib/thesis}

\end{document}

%% file: abs/abstract_en.tex
\begin{abstract}

Large Language Models (LLMs) have attracted considerable attention in recent years due to their remarkable compatibility with Hardware Description Language (HDL) design. In this paper, we examine the performance of three major LLMs, Claude 3 Opus, ChatGPT-4, and ChatGPT-4o, in designing finite state machines (FSMs). By utilizing the instructional content provided by HDLBits, we evaluate the stability, limitations, and potential approaches for improving the success rates of these models. Furthermore, we explore the impact of using the prompt-refining method, To-do-Oriented Prompting (TOP) Patch, on the success rate of these LLM models in various FSM design scenarios. The results show that the systematic format prompt method and the novel prompt refinement method have the potential to be applied to other domains beyond HDL design automation, considering its possible integration with other prompt engineering techniques in the future.

\end{abstract}

%% file: chapters/1-RISCV_SRAM.tex
\section{Introduction}  

\noindent The emergence of Large Language Models (LLMs) has transformed various domains, with their potential applications expanding into new areas. One such field of interest is the automation of chip design, where LLMs have exhibited remarkable compatibility with Hardware Description Language (HDL) design. Mainstream LLMs, such as ChatGPT and Claude, have showcased their capability to aid in HDL design automation. 

Prior research has investigated the use of LLMs for HDL design using natural language, identifying several effective prompt engineering (PE) techniques~\cite{lu2024rtllm} \cite{hammond2020dave}. Extending this groundwork, our study aims to compare the abilities of various LLMs and propose strategies for enhancing their performance. Drawing inspiration from recent evaluation frameworks commonly employed in the field \cite{liu2023verilogeval} \cite{thakur2023autochip}, we utilize a set of problems sourced from the Verilog learning platform, HDLBits, as our test problem set to assess the models' capabilities.

The primary objective of this study is to evaluate the performance of mainstream LLMs, specifically Claude 3 Opus, ChatGPT-4, and ChatGPT-4o, in designing finite state machines (FSMs). By leveraging the instructional content provided by HDLBits, we examine the stability, limitations, and potential approaches for improving the success rates of these models.

For this research, we meticulously selected 20 FSM design problems from HDLBits. We employ a systematic format to address the problem description. By endeavoring to solve these problems using the chosen LLMs in conjunction with the systematic format prompt, we record their success rates and analyze the underlying causes of any encountered failures. Our goal is to develop and propose strategies that can boost the performance of these models in HDL design tasks.

Another key focus of this paper is to explore the impact of using the prompt-refining method, To-do-Oriented Prompting (TOP) Patch. This is an efficient approach to refine the systematic format prompt and achieve better performance. By examining the impact of the TOP Patch on the success rates of LLMs in various FSM design scenarios, we gain insights into the effectiveness of this approach. Furthermore, we explore the potential applications of the TOP Patch beyond FSM design, considering its possible integration with other prompt engineering techniques in the future.

Through the comparative analysis, the systematic format prompt, and the novel prompt refinement method, we aim to contribute to the ongoing advancement of LLMs in the field of HDL design automation. By identifying the strengths and weaknesses of each model, we strive to pave the way for more effective and efficient design processes. Ultimately, our goal is to further enhance the capabilities of LLMs in the realm of automatic chip design, enabling them to become invaluable tools for engineers and researchers in the field.
\section{Effective Prompt Engineering Techniques}

\noindent To enhance the LLMs' comprehension of the provided prompt, we employ a systematic approach using the markdown format. This method allows for clear distinctions between each part of the prompt, such as the specification, I/O list, and module functionality. 

By utilizing itemized lists to explain several points, the emphasis becomes more apparent compared to using a long paragraph. Additionally, the convenience of table construction enables the systematic presentation of I/O details, the conversion of visualized waveform into readable lists with signal names and values, and the construction of state transition tables. By leveraging the markdown format, we aim to improve the LLMs' understanding of the given prompt.

Focusing on the FSM design problems selected from HDLBits, the design of the state transition table undoubtedly represents the high-level language description of the FSM design. We preserve the problem description excerpted from the HDLBits website and transform the example waveform and state description graph into markdown table form. The prompt begins with the command "Act as a professional SystemVerilog programmer." We then place the problem description in the "Specification" section, followed by the "Example waveform" and "Module declaration" sections, where we include the module name and I/O ports. The final part of the prompt asks the LLMs to implement the FSM in SystemVerilog. See an example in \reflong{Prompt}{prompt:1}.

\begin{prompt}[Systematic markdown form of Lemmings1\label{prompt:1}]
    Act as a professional SystemVerilog programmer. You are going to design a module based on the specifications.
\smallskip

\#\#\# Specification

The game Lemmings involves critters with fairly simple brains. So simple that we are going to model it using a finite state machine.

... (remaining problem description)
\smallskip

\#\#\# Example Behavior (truncated)
\begin{verbatim}
| clk        | 1 | 2 | 3 | 4 | 5 |
|------------|---|---|---|---|---|
| bump_left  | 0 | 0 | 0 | 0 | 0 | ...
| bump_right | 0 | 0 | 1 | 0 | 0 |
| walk_left  | 1 | 1 | 1 | 1 | 1 |
| walk_right | 0 | 0 | 0 | 0 | 0 |
\end{verbatim}
\smallskip

\#\#\# Module Declaration

```systemverilog

module top\_module(

    input clk,
    
    input areset,    // Freshly brainwashed Lemmings walk left.
    
    input bump\_left,
    
    input bump\_right,
    
    output logic walk\_left,
    
    output logic walk\_right

);

...

endmodule

```
\end{prompt}

\section{Evaluating the Performance of LLMs in SystemVerilog FSM Design}

\subsection{Experimental Setup and Methodology}

\noindent For the 20 FSM design problems selected from HDLBits, we evaluate each problem five times (using five independent chat sessions) through Claude 3 Opus, ChatGPT-4, and ChatGPT-4o, requiring them to generate SystemVerilog code. 

The problems are nearly identical to those on the HDLBits website, with minor modifications to better suit the markdown format. No additional hints about the problems are provided to the LLMs. For problems that require information from the previous problem, we additionally supply the correct version of the transition table or SystemVerilog code generated by the LLM itself in the preceding problem. Furthermore, to test their SystemVerilog writing capabilities, only the module interface is given. This test aims to assess the basic syntax knowledge and FSM design ability of the LLMs.

\subsection{Comparative Analysis of LLM Performance}

\begin{table}[!ht]
  \begin{center}
    \begin{tabularx}{\linewidth}{L|ccc}
      \toprule
      \textbf{Problem} & \textbf{Claude 3 Opus} & \textbf{ChatGPT-4} & \textbf{ChatGPT-4o} \\
      \midrule
      Fsm1                 & \Five  & \Five  & \Four  \\
      Fsm1s                & \Five  & \Four  & \Five  \\
      Fsm2                 & \Five  & \Three & \Four  \\
      Fsm2s                & \Four  & \Five  & \Two   \\
      Fsm3comb             & \Five  & \Five  & \Five  \\
      Fsm3onehot           & \Zero  & \Zero  & \Zero  \\
      Fsm3                 & \Five  & \Four  & \Four  \\
      Fsm3s                & \Five  & \Three & \One   \\
      \scriptsize{Exams/ece241 2013 q4}
                           & \Zero  & \Zero  & \Zero  \\
      Lemmings1            & \One   & \Zero  & \One   \\
      Lemmings2            & \One   & \Zero  & \Two   \\
      Lemmings3            & \Four  & \Three & \Three \\
      Lemmings4            & \Zero  & \Zero & \Zero  \\
      Fsm onehot           & \Zero  & \Zero & \Zero  \\
      Fsm ps2              & \Three & \Zero & \Zero  \\
      Fsm ps2data          & \Zero  & \Zero & \Zero \\
      Fsm serial           & \Zero  & \Zero & \Zero \\
      Fsm serialdata       & \Zero  & \Zero & \Zero \\
      Fsm serialdp         & \Zero  & \Zero & \Zero \\
      Fsm hdlc             & \Zero  & \Zero & \Zero \\
      \midrule
      \scriptsize{\# of Problem Solved Successfully} & 11 & 8 & 10 \\
      Total Success Rate    & 41\% & 32\% & 31\% \\
      \bottomrule
    \end{tabularx}
    \caption{Assessing the Single-Shot Capabilities of LLMs in Automating FSM Design: Results from 5 Independent Trials}
    \label{table:comp}
  \end{center}
\end{table}

\noindent The results are presented in \reflong{Table}{table:comp}, showcasing the strengths and weaknesses of these models when confronted with various FSM design problems. The common syntax errors include multi-driver issues in multiple always blocks, missing or improperly positioned logic declarations, and confusion between \verb|begin...end| and brackets.

One type of problem that poses a challenge is the one-hot design, which requires deriving equations through inspection, i.e., directly forming Boolean equations. All three models perform poorly in this regard. Across all questions, they generally struggle with handling truth tables or Karnaugh maps. While they correctly generate most logic equations, they often make a few mistakes.

For more intricate designs with lengthy descriptions, the models often misinterpret the state-transition logic relationship. Furthermore, the number of states influences the likelihood of FSM correctness. If fewer states are used to form the FSM, individual states might contain unusual and convoluted logic, which is a primary factor leading to failures.

\emph{Claude 3 Opus} achieves the highest success rate among the three models. It demonstrates stability in solving problems within its capabilities and exhibits no syntax error.

\emph{ChatGPT-4} effectively comprehends and processes the provided information. It generally performs well but occasionally makes mistakes that deviate from the intended solution. Its stability is relatively less consistent compared to the other models.

\emph{ChatGPT-4o} performs similarly to ChatGPT-4 and even surpasses it in terms of correctness and stability. However, it is noteworthy that it defaults to treating designs as having asynchronous reset. When dealing with designs that require synchronous reset (such as Fsm2s \& Fsm3s in \reflong{Table}{table:comp}), it often fails to adhere to the prompt. To cope with this defect, we utilize a special refinement strategy that will be further mentioned in Section IV.

\section{TOP Patch: Overcoming LLM Limitations}  
\noindent Even with systematic markdown format prompt engineering, more complex designs may require more sophisticated design descriptions, which can hinder the effectiveness of the received feedback. To address this issue, we need to provide additional hints on important signals or pivotal concepts to help LLMs focus on the key points and clarify essential concepts. Here, we introduce an effective strategy to improve the performance of LLM register transfer level (RTL) code generation: To-do-Oriented Prompting (TOP) Patch.

TOP Patch is an efficient method to solve problems resulting from complex specifications or concepts. It is an extra snippet that can be appended to the end of the systematic markdown format prompt, boosting the quality of LLMs' responses and consequently increasing the success rate of complex RTL design.

TOP Patch introduces a "To-do" section that is added to the end of the prompt, assisting LLMs in concentrating on specific points. The contents of this section are primarily composed of essential signal specifications, constraints, or important concepts, generated either by human experts or LLMs themselves, and presented point by point. LLMs tend to execute these points in order, making the sequence of the contents a crucial factor. Examples of a TOP Patch utilized in solving particular problems can be found in \reflong{Prompt}{prompt:2} and \reflong{Prompt}{prompt:3}.

\begin{prompt}[TOP Patch for Synchronous Reset FSM design\label{prompt:2}]
  \#\#\# To-do
  \begin{enumerate}[label=\arabic*.]
    \item Explain synchronous reset and give a basic example.
    \item Tell the difference between synchronous and asynchronous reset design in SystemVerilog implementation.
    \item Implement the above design specifications in SystemVerilog. 
  \end{enumerate}
\end{prompt}

\begin{prompt}[TOP Patch for One-hot FSM design\label{prompt:3}]
  \#\#\# To-do
  \begin{enumerate}[label=\arabic*.]
    \item Explain "derive equations by inspection".
    \item List out every situation that will result in each next\_state.
    \item Implement the entire SystemVerilog module for the state machine using the above two results.
  \end{enumerate}
\end{prompt}

By utilizing the TOP Patch strategy, we can address the limitations of certain LLMs when dealing with unique or challenging scenarios. The approach involves minimal human intervention, making it an efficient and practical solution. The inclusion of a TOP Patch has been shown to significantly improve the success rate of LLMs in handling these special cases.

\section{Experimental Results} 

\subsection{Addressing Synchronous Reset Issues in ChatGPT-4o with TOP Patch}
\noindent Through observation, we have discovered that ChatGPT-4o struggles to solve the synchronous reset FSM design problems in \reflong{Table}{table:comp}, specifically Fsm2s and Fsm3s. 

The primary issue is that ChatGPT-4o consistently utilizes asynchronous flip-flops to implement the FSM when prompted solely with the plain problem description.

In \reflong{Prompt}{prompt:3}, the first two points serve to ensure that the LLM understands the distinction between synchronous and asynchronous reset and possesses the knowledge to correctly implement both types of flip-flops.

By incorporating this additional verification step, the TOP Patch successfully increases the success rate from 30\% to 70\%, as demonstrated in \reflong{Table}{table:syncr}. This result showcases the capability of the TOP Patch in assisting LLMs to overcome specific design limitations.

\begin{table}[!ht]
  \begin{center}
    \begin{tabularx}{\linewidth}{L|Cc}
      \toprule
      \textbf{Problem} & \textbf{ChatGPT-4o} & \textbf{ChatGPT-4o + TOP Patch}\\
      \midrule
      Fsm2s           & \Two  & \Four  \\
      Fsm3s           & \One  & \Three  \\
      \midrule
      Success Rate    & 30\% & 70\%  \\
      \bottomrule
    \end{tabularx}
    \caption{Single-Shot Capabilities of ChatGPT-4o utilizing TOP Patch in Synchronous Reset FSM Design}
    \label{table:syncr}
  \end{center}
\end{table}

\subsection{Improving One-hot FSM Design Success Rates using TOP Patch}
\noindent For the one-hot FSM design problems in \reflong{Table}{table:comp}, namely Fsm3onehot and Fsm onehot, the design requirements involve deriving equations through inspection, which restricts the design method. 

If prompted solely by the plain problem description, LLMs tend to either ignore or struggle with the complex request. Consequently, specific hints are provided using the TOP Patch approach.

In \reflong{Prompt}{prompt:2}, the first two points aim to draw the LLMs' attention to the pivotal elements that significantly affect correctness. The first point ensures that LLMs fully comprehend and understand the special requirements of this problem. The second point further requests LLMs to summarize every transition situation. By providing these hints, we ensure an increase in the accuracy and success rate of these unique one-hot FSM designs.

After applying the TOP Patch, a noticeable improvement in the success rate is observed. As shown in \reflong{Table}{table:oneh}, the success rates of all three LLMs demonstrate the ability for them to solve the one-hot FSM design problems. Impressively, Claude 3 Opus demonstrates the best performance, achieving a 90\% success rate when using the TOP Patch.

\begin{table}[!ht]
  \begin{center}
    \begin{tabularx}{\linewidth}{L|cCC}
      \toprule
      \textbf{Problem} & \textbf{Claude 3 Opus} & \textbf{ChatGPT-4} & \textbf{ChatGPT-4o}\\
      \midrule
      Fsm3onehot           & \Four  & \Two  & \One  \\
      Fsm onehot           & \Five  & \One & \Three  \\
      \midrule
      Success Rate    & 90\% & 30\% & 40\% \\
      \bottomrule
    \end{tabularx}
    \caption{Single-Shot Capabilities of LLMs utilizing TOP Patch in One-hot FSM Design}
    \label{table:oneh}
  \end{center}
\end{table}

\subsection{Tackling Complex FSM Designs with TOP Patch and Chain-of-Thought Prompting}
\noindent For much more complex problems, it seems impossible to complete them in a single attempt. It may require feedback to revise the error points or inspiration from manual hints. 

In this case, the TOP Patch strategy can also work as a medium to emphasize the tasks to be executed in the next chat section. It provides a workflow and can act as a framework integrated with Chain-of-Thought (CoT), one of the famous PE techniques. \reflong{Prompt}{prompt:fsm} shows an example for Lemmings1 in HDLBits that uses TOP Patch to provide a CoT flow. 

\begin{prompt}[Multi-Shot using TOP Patch \& CoT for Lemmings1\label{prompt:fsm}]
  \#\#\# To-do
  \begin{enumerate}[label=\arabic*.]
    \item Read the example behavior first carefully. Elaborate the reasoning behind these behaviors.
    \item design a FSM for the above specification. Give the state transition table with the outputs details that contains every input condition for each state.
    \item implement the entire SystemVerilog module for the state machine.
  \end{enumerate}
\end{prompt}

Furthermore, in cases where a single task may contain a large amount of response, it is acceptable and potentially beneficial to break it into several prompts. This approach ensures and checks the quality of the current stage, allowing the remaining stages to be gradually completed.

After the initial SystemVerilog codes are generated, some advanced feedback methods can be introduced~\cite{blocklove2023chip}.
Based on the severity of the error, error feedback (directly copying the compilation info), human feedback (manually checking the problems that occurred in the waveform), and advanced feedback (engaging in conversation with LLMs to obtain advice) take place in the secondary debugging step.

\section{Conclusion and Future Directions}
\noindent This research provides a comparison between three major LLMs in terms of their HDL generation quality. Utilizing a problem set from HDLBits, we employ a systematic and effective prompt method and propose an efficient prompt refinement approach called TOP Patch. This refinement method yields impressive results in performance enhancement.

Moreover, the generation of TOP Patch can be accomplished by LLMs that require further training and fine-tuning. In the future, once TOP Patch automation becomes stable, it has the potential to be integrated into the feedback system of automated HDL generator frameworks~\cite{thakur2023autochip}. By incorporating this prompt refinement solution, we can achieve a higher level of chip design automation.

Looking ahead, the automation of TOP Patch generation through fine-tuned LLMs opens up exciting possibilities for enhancing the efficiency and accuracy of HDL design automation frameworks. By seamlessly integrating TOP Patch into the feedback loop of these frameworks, we can create a more robust and adaptive system that continuously refines prompts based on the specific design context and requirements.

Furthermore, the principles and techniques introduced in this research have the potential to be applied to other domains beyond HDL design automation. The systematic prompt method and the concept of prompt refinement through TOP Patch can be explored in various fields where LLMs are utilized for code generation, natural language processing, or creative writing tasks.